\documentclass[a4paper,fleqn,usenatbib]{mnras}

\usepackage{mathptmx}

\usepackage[T1]{fontenc}
\usepackage{ae,aecompl}

\usepackage{graphicx}	
\usepackage{amsmath}	
\usepackage{amssymb}	
\usepackage{BibDef}
\usepackage{geometry}
\usepackage{lipsum}

\newcommand\msun{\, \rm M_\odot}
\newcommand\pc{{\, \rm pc}}
\newcommand\mbin{{M_{\rm b}}}
\newcommand\mgal{{M_{\rm g}}}
\newcommand\nlc{{N_{\rm lc}}}

\title[Brownian motion of massive black hole binaries]{Brownian motion of massive black hole binaries and the final parsec problem}

\author[E. Bortolas et al.]{
E. Bortolas,$^{1,2}$\thanks{E-mail: elisa.bortolas@oapd.inaf.it}
A. Gualandris,$^{3}$
M. Dotti,$^{4,5}$
M. Spera,$^{1}$
M. Mapelli,$^{1,5}$
\\
$^{1}$INAF, Osservatorio Astronomico di Padova, Vicolo dell'Osservatorio 5, I-35122, Padova, Italy\\
$^{2}$Dipartimento di Fisica e Astronomia ''Galileo Galilei'', Universit\`a di Padova, Vicolo dell'Osservatorio 3, 35122 Padova, Italy\\
$^{3}$Department of Physics, University of Surrey, Guildford GU2 7XH, United Kingdom\\
$^{4}$Dipartimento di Fisica G. Occhialini, Universit\`a degli Studi di Milano, Bicocca, Piazza della Scienza 3, 20126 Milano, Italy\\
$^{5}$INFN, Sezione di Milano-Bicocca, Piazza della Scienza 3, 20126 Milano, Italy\\
}

\date{Accepted XXX. Received YYY; in original form ZZZ}

\pubyear{2016}

\begin{document}
\label{firstpage}
\pagerange{\pageref{firstpage}--\pageref{lastpage}}
\maketitle

\begin{abstract}
 Massive black hole binaries (BHBs) are expected to be one of the most powerful sources of gravitational waves (GWs) in the frequency range of the pulsar timing array and of forthcoming space-borne detectors. They are believed to form in the final stages of galaxy mergers, and then harden by slingshot ejections of passing stars. 
 However, evolution via
  the slingshot mechanism may be ineffective if the reservoir of
  interacting stars is not readily replenished, and the binary
  shrinking may come to a halt at roughly a parsec separation.  Recent
  simulations suggest that the departure from spherical symmetry,
  naturally produced in merger remnants, leads to efficient loss cone
  refilling, preventing the binary from stalling.  However, current
  $N$-body simulations able to accurately follow the evolution of BHBs
  are limited to very modest particle numbers. 
Brownian motion may artificially  enhance the loss cone refilling rate in low-$N$ simulations, where
  the binary encounters a larger population of stars due its random
  motion. Here we study the
  significance of Brownian motion of BHBs in merger remnants in the
  context of the final parsec problem. We simulate mergers with various particle numbers (from 8k to 1M) and with several density profiles. Moreover, we compare simulations where the BHB is fixed at the centre of the merger remnant with simulations where the BHB is free to random walk.
We find that Brownian motion does not significantly affect
  the evolution of BHBs in simulations with particle numbers in excess
  of one million, and that the hardening measured in merger
  simulations is due to collisionless loss cone refilling. 
\end{abstract}



\section{Introduction}

Astrophysical observations suggest that massive black holes (MBHs)
with masses in the range $10^6-10^{10}\msun$ inhabit a large fraction
of galactic nuclei, and their presence in most massive galaxies is
suspected to be ubiquitous { \citep[e.g.][]{ff05}}.  In
addition, according to the $\Lambda$CDM cosmological model, galaxies
grow through the agglomeration of smaller structures, at least some of
which contain a black hole seed at early times\footnote{As can be inferred by the fact that galaxies turn on as active galactic nuclei for a small fraction of their lifespan \citep{haehneltrees}.}.  Therefore BHBs may form in large numbers along cosmic
time, as a result of galaxy mergers.

The formation and evolution of BHBs has received considerable
attention in the last years, as the merger and final coalescence of
MBH pairs is expected to represent the loudest source of GWs in the $10^{-4}-10^{-1}$ Hz band \citep{gw}. These
low-frequency signals cannot be observed by ground-based
interferometers like LIGO and VIRGO, which have recently detected the
first ever GW waveform from the coalescence of two stellar mass black
holes \citep{Abbott2016}. However, the lower mass MBH mergers will be
accessible to the new generation of space-based gravity
interferometers, such as \emph{eLISA}, the straw-man mission of
\emph{The Gravitational Universe} theme, selected by \emph{ESA} for L3
\citep{elisa}.  The most massive BHBs, on the other hand, can be
detected by the Pulsar Timing Array \citep{Babak2016} which is
already operational and using the very accurate timings of pulsars to
observe GWs. The detection of low-frequency GWs from MBHs would
provide crucial information on the masses and spins of MBHs and on the
merger rate of BHBs, which would allow to constrain models of their
formation and growth \citep[][]{volonteri10}.  
However, the coalescence of BHBs has been put under scrutiny in the past years due to theoretical arguments on loss cone refilling \citep[e.g.][]{begelman} and $N$-body simulations of
BHB pairing in isolated spherical galaxies
\citep[e.g.][]{makinofunato}, both hinting at a possible stalling
of the BHBs at $\sim{}1$ pc scales, an issue commonly referred to as the
``Final Parsec Problem''.

The path to coalescence of a BHB in gas-poor galaxies can be divided
into three main phases \citep{begelman}: (i) a first phase during
which dynamical friction drags the MBHs toward the centre of the
merger remnant; (ii) a second phase in which subsequent binary
shrinking is induced by three-body interactions between the
BHB and the surrounding stars (the so-called \emph{slingshot
  mechanism}), leading to the ejection of stars from the centre of the
remnant \citep{slingshot}; (iii) a third phase of evolution
characterised by either stalling at roughly a parsec separation or
inspiral due to {GW} emission followed by coalescence to a single
MBH. While the transition from the dynamical friction to the slingshot
phase is quick for any realistic MBH masses, the efficiency of
gravitational slingshot interactions drops after roughly a dynamical
time, when the population of stars on initially low-angular momentum
orbits has been ejected by the binary, possibly leading to
stalling. The fate of the BHB at this stage depends crucially on the
supply of stars in the \emph{loss cone}, i.e. the low angular momentum
region of the phase space harbouring stars that can experience an
interaction with the binary \citep{begelman,yu,milosav}.

In principle, two-body relaxation always contributes to loss cone
refilling via star scatterings, however it operates on a timescale that
strongly depends on the number $N$ of particles in the system (roughly
$T_{\rm r} \propto N / \log N$), and this exceeds the age of the
Universe  in almost all { sufficiently luminous galaxies (i.e. galaxies harbouring MBHs with masses in excess of $10^7 \msun$).}  
 { Consequently, relaxation-induced loss cone refilling is generally assumed to
be ineffective,
 }
 and the
binary would not be able to reach the GW inspiral phase in a Hubble
time.  Direct-summation $N$-body simulations of BHB inspiral in
spherical stellar environments confirm the stalling and the dependence
on particle number \citep{makinofunato,BMS2005,merritt2007}.  Simulations
show a completely different behaviour, though, when the merger of two
galaxies hosting a central MBH is followed from early times
\citep{preto,khan11,gualandrismerritt}. BHBs formed during mergers are
able to continue hardening well after the first central depletion,
which implies that the loss cone is efficiently replenished at all
times. Because merger remnants are typically non-spherical, the
sustained binary hardening is attributed to the population of
centrophilic orbits unique to  non-spherical potentials. In these cases, BHB
coalescence can be reached in a few Gyr at the most.  Collisionless
loss cone refilling has been confirmed as the driver of BHB evolution
in simulations of binaries evolving in isolated triaxial galaxies
\citep[][Gualandris et al., in prep]{vasiliev15}, in which
collisional refilling is reduced or eliminated by using a different
numerical technique and increasing particle number significantly.

However, particle numbers representative of real galaxies remain
unachievable by $N$-body simulations and, unless collisional
relaxation is eliminated in some way, loss cone refilling due to
two-body relaxation cannot be avoided. This is particularly true for
direct-summation methods, which are needed when one wishes to model
the detailed evolution of the BHB, but whose $N^2$ computational
complexity limits applicability to a few million particles at most.

{ Simulations with low $N$} might be plagued by an increased
Brownian motion of the binary's centre of mass (CoM). Whenever a
slingshot interaction occurs, the binary CoM experiences a recoil
whose amplitude depends on the mass of the ejected star. In low-$N$
models, where stars are orders of magnitude more massive than in real
galaxies, the binary experiences an enhanced Brownian motion.  As it
wanders within its typical Brownian radius, the binary can intersect
more stars than it would were it confined to a smaller
region. Therefore, the loss cone is artificially enhanced.

{ What is the importance of Brownian motion and whether it depends on the resolution has been debated for a long time. It has been argued that \emph{``a substantial fraction of all massive binaries in galaxies can coalesce within a Hubble time''} due to the effects of Brownian motion \citep{chatterjee}. In contrast, other theoretical studies suggest that Brownian motion is modest for BHBs in real galactic nuclei \citep{merrittbrown01, milosav},} but it remains to
be established whether it is responsible for the sustained hardening
of BHBs observed in direct-summation merger simulations.  Here we
study the Brownian motion of BHBs in merger simulations and its
connection with loss cone repopulation and the final parsec problem.

We performed a suite of direct-summation $N$-body simulations of the
merger of two spherical gas-free galaxies containing a MBH in the
centre. We measure the wandering of the BHB that forms and its
hardening rate and show that the Brownian motion does not
significantly affect the binary evolution at large particle numbers.
In order to draw a more robust conclusion, we perform additional
simulations in which we artificially anchor the binary's CoM to the
centre of the density cusp as the system evolves. We find that the
binary shrinking rate in these simulations is in good agreement with
the one found in the free-binary case when the resolution
(i.e. particle number) of our simulations is maximal. This allow us to
completely rule out the possibility of a spurious wandering-induced
loss cone refilling in simulations with $N$ in excess of one million.

The paper is organised as follows: section \ref{sec:methods} details
the numerical methods and initial conditions adopted; section
\ref{sec:sim} presents results of the BHB evolution, while we discuss
Brownian motion and its dependence on resolution in section
\ref{sec:bm}; finally, in section \ref{sec:sum} we present a summary
and conclusions.

\section{Methods}
\label{sec:methods}
We performed a suite of direct-summation $N$-body simulations of equal
mass galaxy mergers, varying the resolution (i.e. particle number $N$)
and the density distribution of the progenitor galaxies. Each galaxy
follows a Dehnen's density profile \citep{dehnen}:
\begin{equation}
\rho(r)=\frac{(3-\gamma) \,\mgal}{4\pi}\frac{r_0}{r^\gamma(r+r_0)^{4-\gamma}},
\end{equation}
with total mass $\mgal$, scale radius $r_0$ and inner slope $\gamma$.
Two different realisations of the same model were placed on a bound
elliptical orbit in the $x-y$ plane, with initial orbital eccentricity
$e = 0.5$, separation $\Delta r = 20r_0$ and semimajor axis $a =
15r_0$.  The units were chosen in such a way that
$G=r_0=M_{\mbox{tot}}=1$, where $M_{\mbox{tot}}$ represents the total
stellar mass involved in each merger simulation. A point mass
representing the MBH was placed at the centre of each spherical
system, with a mass $M_{\bullet} = 0.005 \mgal$.

\begin{table}
  \centering
  \caption{List of {simulations.  Column~1: simulation set (A, B, C); column~2: run name; column~3: total number of particles $N$; column~4: slope $\gamma$ of the inner profile in each galaxy model.} All mergers are equal mass.}
  \label{tab:ic}
  \begin{tabular}{llccc} 
    \hline
 Set &   Run & $N$ & $\gamma$ \\
    \hline
 A   & 1 & 8k   & 0.5 \\
     & 2 & 16k  & 0.5 \\
     & 3 & 32k  & 0.5 \\
     & 4 & 64k  & 0.5 \\
     & 5 & 128k & 0.5 \\
     & 6 & 256k & 0.5 \\
     & 7 & 512k & 0.5 \\
     & 8 & 1M   & 0.5 \\
    \hline
B    & 1 & 8k   & 1.0 \\
     & 2 & 16k  & 1.0 \\
     & 3 & 32k  & 1.0 \\
     & 4 & 64k  & 1.0 \\
     & 5 & 128k & 1.0 \\
     & 6 & 256k & 1.0 \\
     & 7 & 512k & 1.0 \\
     & 8 & 1M   & 1.0 \\
    \hline
C    & 1 & 8k   & 1.5 \\
     & 2 & 16k  & 1.5 \\
     & 3 & 32k  & 1.5 \\
     & 4 & 64k  & 1.5 \\
     & 5 & 128k & 1.5 \\
     & 6 & 256k & 1.5 \\
     & 7 & 512k & 1.5 \\
    \hline
  \end{tabular}
\end{table}

We performed three main groups of simulations, each characterised by a
different inner slope for the colliding galaxies: $\gamma=0.5$ for { set~A}, $\gamma=1$ for { set~B,} and $\gamma=1.5$ for { set C}. For each
set, we considered different values for $N$ from 8k to 1M. For set
C the $N$=1M case was avoided due to the prohibitively long
integration time required. The list of performed simulations is given
in Table \ref{tab:ic}.  Each model was evolved for $t \approx 500$
time units; this choice ensures that the binary is followed for
$\Delta t \gtrsim 100$ after the MBHs form a bound Keplerian pair.   The parameters of the performed $N$-body
experiments are summarised in Table \ref{tab:pars}.

The initial conditions were evolved adopting the integrator
\texttt{HiGPUs}, a direct-summation $N$-body code based on the sixth
order Hermite scheme with block timesteps, designed to run on GPUs
\citep{higpus}. The integrator adopts a Plummer softening kernel
\citep{plummer}; we chose a softening parameter $\varepsilon=10^{-4}$;
this length scale is smaller than the minimum separation reached by
the binary during its evolution.  In \texttt{HiGPUs}, the individual
particle timesteps are computed via a combination of the sixth and 
fourth order Aarseth criterion \citep{aarsethbook,nitadorimakino}, and
the two accuracy parameters were set equal to $\eta_{\rm sixth}=0.45$,
$\eta_{\rm fourth}=0.01$ \citep[for details, see][]{higpus}.  The
minimum and maximum possible values in the hierarchy of timesteps were
chosen as $\Delta t_{\rm min}=2^{-29}\approx 1.863\cdot 10^{-9}$ and $\Delta
t_{\rm max}=2^{-6}=0.015625$.

 All simulations of set A, B, and C were run allowing the BHB to wander (hereafter, \emph{free-binary simulations}). Simulations of set B were then re-run while periodically re-centering the BHB at the centre of the merger remnant to quench the binary's wandering (hereafter, \emph{fixed-binary simulations}).

\begin{table}
  \centering
  \caption{Parameters of the simulations. From left to right: initial
    semi-major axis, initial eccentricity, initial separation of the
    two original galaxies, scale radius of each galaxy, total
    stellar mass in each simulation, mass ratio between the two
    merging galaxies,  mass ratio between the BHB and the stellar
    mass in the simulation,  softening parameter.}
  \label{tab:pars}
  \begin{tabular}{lccccccccc} 
    \hline
    $a_{\mbox{i}}$ & $e_{\mbox{i}}$ & $\Delta r$ & $r_0$ & $M_{\mbox{tot}}$ & $q$ &$\mbin/M_{\mbox{tot}}$ & $\varepsilon$ \\
    \hline
    \hline
    15 & 0.5 & 20 & 1 & 1 & 1 & 0.005 &$10^{-4}$  \\
    \hline
  \end{tabular}
\end{table}

\section{Simulations and binary evolution}
\label{sec:sim}
\begin{figure}
\includegraphics[angle=270,trim={9.5cm 0cm 1.2cm 0cm},width=\columnwidth]{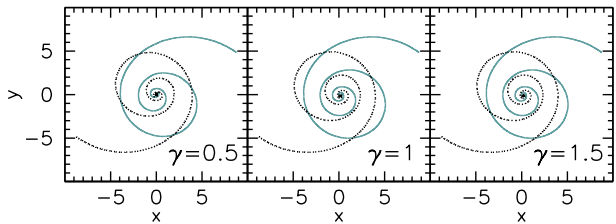} 
    \caption{Trajectories of the two MBHs in the $x-y$ plane for the
      three sets of simulations, corresponding to $\gamma=0.5,1.0,
      1.5$, from left to right, and $N=512$k.  In this Figure and in the following, distances are in scalable $N-$body units.}
    \label{fig:path}
\end{figure}
The trajectories of the MBHs in the $x-y$ plane are shown in
Figure~\ref{fig:path} for the three simulation sets.  The initial
evolution of the MBH pair reflects the orbit of the progenitor
galaxies, which is the same for all three sets, and the trajectories
are therefore quite similar.

\begin{figure*}
	 \includegraphics[angle=270,trim={9cm 0 2cm 0},width=\textwidth]{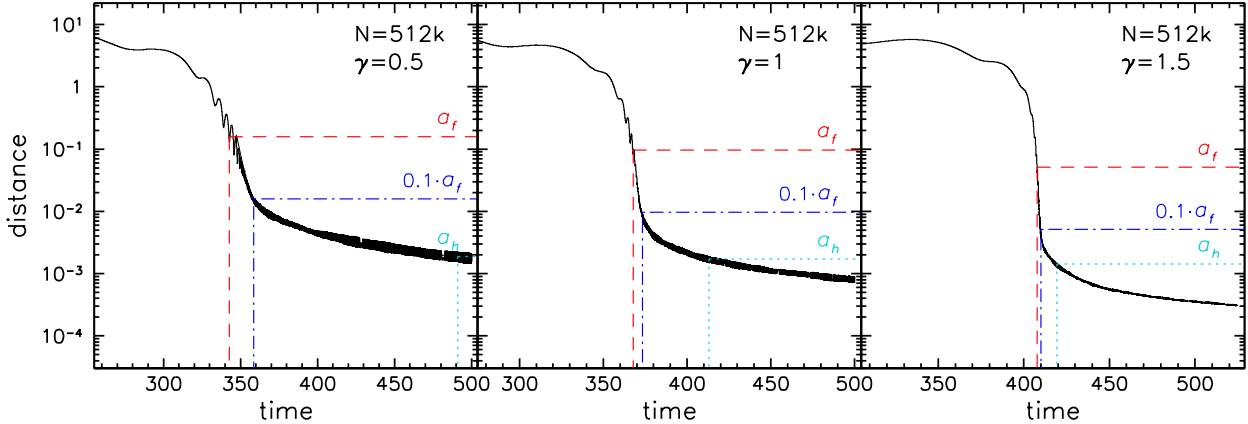}
    \caption{Evolution of the binary separation as a function of time for the runs
      with resolution $N=$512k. From left to right, panels show the
      evolution for models with initial central density slope
      $\gamma=0.5$, $\gamma=1$ and $\gamma=1.5$. The significant
      separations $a_{\rm f}$ (roughly at which the Keplerian binary forms,
      equation \eqref{eq:af}), $a_{\rm f}/10$ and $a_{\rm h}$ (from which slingshot
      interactions expel unbound stars, equation \eqref{eq:ah}) and
      their corresponding times are marked in the panels.  In this Figure and in the following, times and distances are in scalable $N-$body units.}
    \label{fig:binseparation}
\end{figure*}
{ The time evolution of the binary separation is shown in Figure \ref{fig:binseparation}, for the three different sets (free-binary simulations), and for $N=512$k.} This represents the maximum resolution available
for all three $\gamma$ values, and is therefore our fiducial resolution.

We can identify the typical three phases of BHB evolution in all our
simulations.  We consider an equal mass binary with total mass
$\mbin$. The first phase of evolution is driven by dynamical friction,
which brings the {MBH} pair to a separation $a_{\rm f}$, defined as the
separation at which the stellar mass $M_*$ enclosed in the binary
orbit is twice the mass of the secondary MBH. In our case:
\begin{equation} \label{eq:af}
M_*(a_{\rm f})= \mbin.
\end{equation}
The time $t(a_{\rm f})$ at which this separation is reached approximately
corresponds to the time of formation of a bound Keplerian binary in
our equal mass simulations.  Around this time, stellar encounters with
stars on intersecting orbits start becoming important, and mark the
beginning of the \emph{strong three-body scattering regime}
\citep{sesa10}. In this phase, which sees a rapid shrinking of the
binary separation, slingshot ejections first combine with dynamical
friction and then become the main driver of binary hardening. This
phase is rather short, and ends when the loss cone consisting of stars
initially on intersecting orbits has been emptied. At this point, the
inner cusp of the remnant is destroyed and a core has been carved in
the stellar distribution.

We define the binary to be hard when its binding energy per unit mass
exceeds the kinetic energy per unit mass of the field stars, i.e. when
it reaches a separation
\begin{equation}\label{eq:ah}
a_{\rm h} =\frac{G\,{}\mbin}{8\sigma_*^2}.
\end{equation}
Here $\sigma_*$ represents the velocity dispersion of the field stars
\citep{milosav2}. From this moment, stars ejected from the nucleus via
the slingshot mechanism attain a velocity greater than the escape
velocity. The quantities $a_{\rm f}$ and $a_{\rm h}$ and their corresponding times
are marked in Figure \ref{fig:binseparation}. The remnant mass
  distribution keeps memory of the initial slope $\gamma$, especially
  before the mass-carving has taken place; as a consequence, both
  $a_{\rm f}$ and $a_{\rm h}$ show a clear dependence on the initial value of
  $\gamma$ (see Figure \ref{fig:binseparation}): the radius containing
  the mass of the binary is smaller for merger products generated by
  steeper initial models, inducing a smaller $a_{\rm f}$, while the velocity
  dispersion $\sigma_*^2$ of cuspier models is expected to be higher
  and produces a smaller $a_{\rm h}$. 

\begin{figure*}
	\includegraphics[angle=270,trim={3cm 0 2cm 0},width=\textwidth]{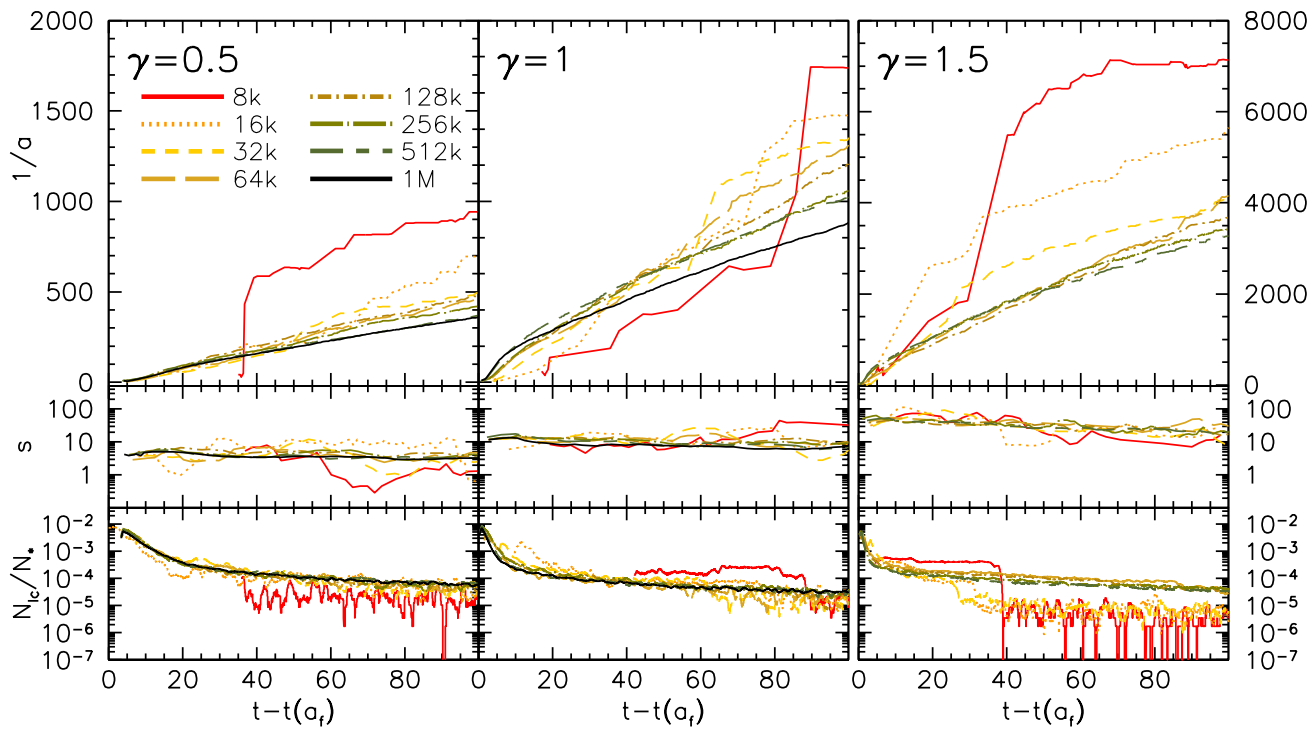} 
        \caption{Time evolution of the inverse semimajor axis $1/a$
          (top panels), hardening rate (middle panels) and fraction of
          loss cone stars (bottom panels) for all the available
          resolutions from $N=8$k to $N=$1M. { We averaged the hardening rate over several time intervals, to smooth fluctuations.}
The { left-hand} panels refer to
          the set with initial density slope $\gamma=0.5$, the central
          panels refer to the set with $\gamma=1$ and the { right-hand} panels
          refer to the last set with $\gamma=1.5$. Notice the
          different scale of the vertical axis in the top right panel.
          Time on the horizontal axis is given from the time
          corresponding to $a = a_{\rm f}$;}
    \label{fig:evolution}
\end{figure*}

We also define the hardening rate 
\begin{equation}
s(t)=\frac{\mbox{d}}{\mbox{d}t}\frac{1}{a},
\end{equation}
where $a$ represents the semi-major axis of the binary.  We compute
$s$ in all simulations by performing a linear fit to $a^{-1}(t)$ in
small time intervals, and plot its time evolution in the middle panels
of Figure \ref{fig:evolution}. The hardening rate is approximately
constant with time.  This behaviour is expected for binaries hardening
in constant density backgrounds \citep{hills}, i.e. when the stellar field is
unaffected by the binary evolution. Since this is not the case in a
merger, the roughly constant nature of $s$ implies that the loss cone
is efficiently replenished, in agreement with previous studies that found the BHB to 
merge within a Hubble time \citep{preto,khan11,gualandrismerritt}.

The Figure also shows the evolution of the inverse semi-major axis
$1/a$ (top panels) and the number of stars inside the loss cone
$\nlc$, normalised to the total number of stars. Since hardening is
assumed to be due to stars in the loss cone interacting with the
binary, it is useful to monitor how many stars can be found in the
loss cone at any given time.  A star is considered to belong to the
loss cone if its angular momentum per unit mass ($L_*$) is smaller
than the angular momentum per unit mass of the binary, i.e. $ L_* \le
L_{\rm b} = \sqrt{G\,{} \mbin \,a(1-e^2)}$. We find that after the binary has
become hard, the fraction of stars in the loss cone is approximately
constant, or only slightly decreasing, on the time scale considered in
the simulations. The decrease is due to the fact that the loss cone
shrinks with time as the binary separation shrinks \citep{vasiliev15}.

\section{Brownian motion}
\label{sec:bm}

\subsection{Theoretical expectations}
The Brownian motion of a single MBH in a nucleus is well described by
a random walk due to encounters with field stars \citep{chandra,dinfric}. If a massive object with mass $M$ is surrounded by
an isothermal distribution of stars, its velocity dispersion $\sigma$
is expected to obey energy equipartition with the background stars:
\begin{equation}\label{eq:bm}
\sigma^2=\frac{m}{M}\sigma^2_*,
\end{equation}
where $\sigma_*$ is the stellar velocity dispersion and $m$ is the
typical stellar mass. This result has been shown to remain valid for a
wide range of stellar backgrounds
\citep{dorband,merritt05,merritt07}. A similar behaviour is expected
for the CoM of a BHB in a galactic nucleus.  \emph{Superelastic
  scattering} and the fact that stars are ejected in random directions
slightly enhance the Brownian motion amplitude in the binary case;
however, \citet{merrittbrown01} showed that the binary's random motion
in phase space is dominated by distant encounters and therefore the
wandering essentially obeys equation \eqref{eq:bm}. \citet{merrittbrown01}
derived an expression for the characteristic amplitude of the motion
$r_{\rm b}$, under the assumptions that (i) the binary is placed in a
constant density core, (ii) the stellar potential is harmonic and (iii)
the Brownian motion has no influence on the stellar velocity field and
mass distribution:
\begin{align}
r_{\rm b} = & \left(\frac{R_2}{R_1}\right)^{1/2} \left(\frac{m}{\mbin}\right)^{1/2} r_{\rm c} \nonumber\\
    = & 0.01\pc \left(\frac{R_2}{R_1}\right)^{1/2} \left(\frac{m}{\msun}\right)^{1/2} 
\left(\frac{10^8\msun}{\mbin}\right)^{1/2} \left( \frac{r_{\rm c}}{100\pc}\right),
\end{align}
where $(R_2/R_1)^{1/2}$ is a factor of order unity related to
Chandrasekhar's coefficients, $r_{\rm c}$ represents the King core radius
\citep{king} and $\mbin$ is the total mass of the binary.

\subsection{Free-binary simulations}
In order to characterise the Brownian motion of the BHB, it is
necessary to define a meaningful reference position to be compared
with the binary CoM over time. In principle, one could think of using
the centre of mass of the whole stellar system, but this is dominated
by the outermost stars which have a negligible influence on the BHB
evolution. A better option is the centre of mass of the 50\% innermost
stars in the system, which allows one to discard the outermost stars
and consider only stars belonging to the density cusp to define the
reference centre.  Since at early times two stellar cusps are present,
we started the Brownian motion characterisation from the moment when
the binary separation reaches $a_{\rm f}/10$: at this
stage, the Keplerian binary has formed and the two original systems
have already merged into a single cusp\footnote{ This particular moment
  has been chosen since it  corresponds approximately to the point when the
  binary decay slows down considerably in all the three sets of
  simulations performed. We avoided other approaches due to the noisy
  trend of the binary separation in the lower-$N$ cases.}. 

We evaluated the typical
wandering radius of the binary in the following way: (i) first we
computed the CoM (C0) from 50\% of the particles closest to the
binary's CoM; (ii) then we reiterated the procedure by evaluating a
more accurate CoM (C1) over the 50\% stars closer to C0, once this
quantity is known. The displacement of C1 from the BHB CoM constitutes
our best estimate of the binary wandering amplitude $r_{\rm b}$ over
time. It is worth stressing that our recursive approach avoids biases
induced by the selection of stars with respect to the position of the
BHB.

The proportionality relation $\sigma^2 \propto m$ from equation
\eqref{eq:bm} can be verified by computing the BHB CoM velocity
dispersion $\sigma^2$ with respect to the CoM velocity of the stellar
background $\sigma_*$. { This} velocity CoM has been computed over the
same stars used for the calculation of C1.

\begin{figure*}
  \includegraphics[angle=270,trim={1cm 0 2cm 0},width=\textwidth]{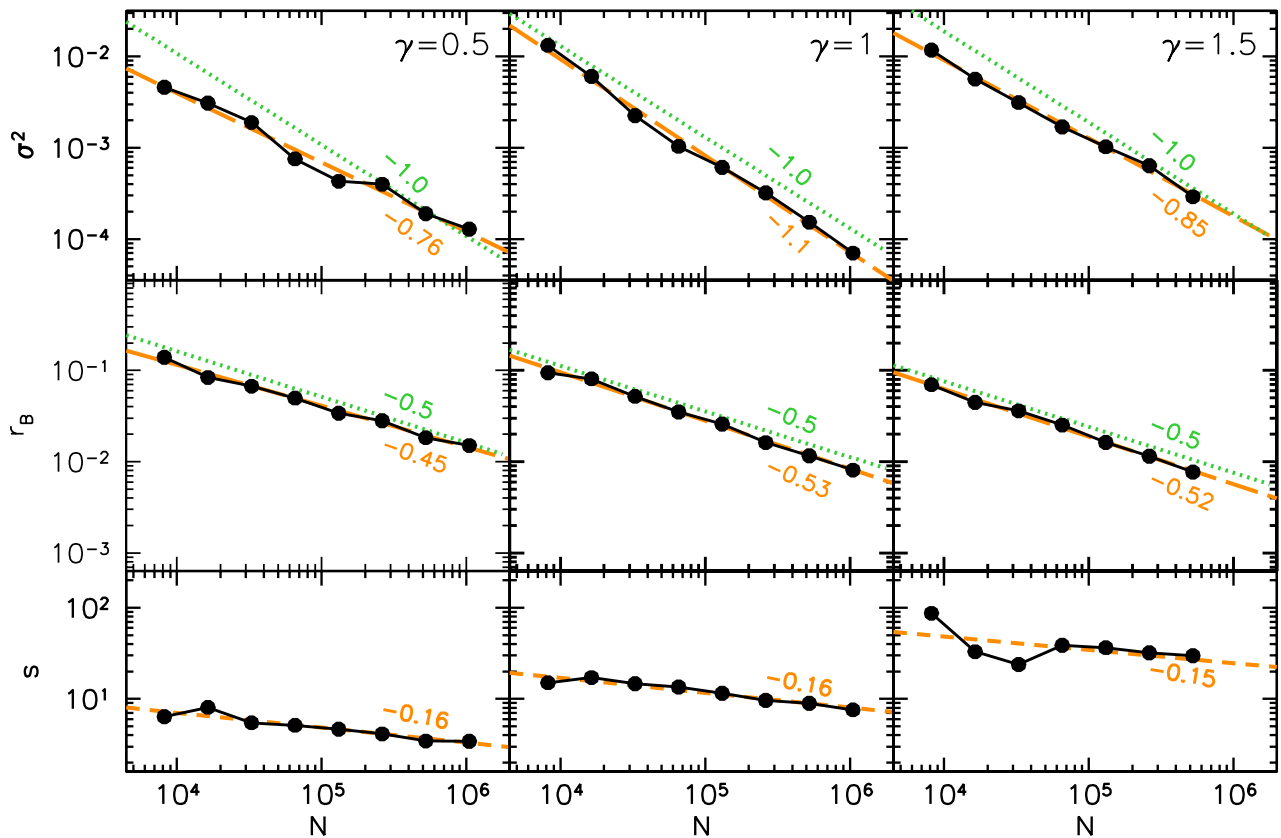}
  \caption{Analysis of the BHB Browninan motion. Top: Time-averaged
    binary velocity dispersion as a function of the number of
    particles $N$ in each simulation. Centre: Time-averaged wandering
    radius of the BHB as a function of particle number.  Bottom:
    Time-averaged hardening rate versus particle number.  In all
    cases, the average is computed from the time when a separation
    $a_{\rm f}/10$ is reached and over 100 time units thereafter. The orange
    dashed lines represent power-law fits to the data and the obtained
    best-fit slopes are given by the orange labels below the lines.
    The green dotted lines indicate the expected dependencies based on
    theoretical models, with slopes given by the green labels above
    the lines.  In all cases, the { left-hand} panels refer to simulations
    with $\gamma=0.5$, central panels refer to simulations with
    $\gamma=1$, and { right-hand} panels to the set with $\gamma=1.5$.}
    \label{fig:bm}
\end{figure*}

In order to test the binary Brownian motion dependence on the number
of particles in each simulation, we averaged the hardening rate $s$,
binary wandering radius $r_{\rm b}$ and CoM velocity dispersion $\sigma^2$
over time from the moment at which the binary separation reaches
$a_{\rm f}/10$ and for $\Delta t$=100 time
units.  From equation \eqref{eq:bm}, we expect a theoretical dependence of
$r_{\rm b} \propto N^{-0.5}$ and $\sigma^2\propto N^{-1}$, given that
$m=1/N$ in the simulations.  The averaged $\sigma^2$, $ r_{\rm b}$ and $s$
are shown in Figure \ref{fig:bm} as a function of $N$.
We performed a power-law fit of the obtained data points for each of
these quantities as a linear least-square fit in logarithmic scale.
We thus obtained a dependence of the form $r_{\rm b}\propto N^{p_r}$,
$\sigma^2\propto N^{p_{\sigma^2}}$ and $s\propto N^{p_s}$. The
computed power-law indexes are listed in Table \ref{tab:bm} and show a
good agreement with the expected values $p_{\sigma^2}=-1$ and
$p_{r}=-0.5$.  Therefore the observed amplitude of the binary's
Brownian motion follows the expected theoretical dependence on
particle number. 

If the loss cone refilling were driven by Brownian
motion, a similar dependence on $N$ would be expected for the
hardening rate. However, this is not observed, and the hardening rate
of the binary shows a much weaker dependence on $N$ than the amplitude
of the wandering. We argue that such dependence is due to a
combination of collisional repopulation of the loss cone and Brownian
motion at these moderate values of $N$, but should become negligible
at more realistic particle numbers (Gualandris et al., in
preparation).
\begin{table}
	\centering
	\caption{List of best-fit power-law indexes of the BHB CoM
          velocity dispersion ($\sigma^2$), wandering amplitude
          ($r_{\rm b}$) and hardening rate ($s$). { Column~1: index name; column~2: expected values based on theoretical arguments; columns~3--5: results for simulations whose initial
          galaxies had inner slope $\gamma=0.5, 1.0, 1.5$, respectively}. }
	\label{tab:bm}
		\begin{tabular}{lrccc} 
		\hline
		 Index& exp. & $\gamma=0.5 $ & $\gamma=1$ & $\gamma=1.5$\\
		 \hline
		 $p_{\sigma^2}$&$-$1&$-$0.76&$-$1.1&$-$0.86 \\
		 
		 $p_{r_{\rm b}}$&$-$0.5&$-$0.45&$-$0.53&$-$0.52\\
		 
		 $p_s$&0&$-$0.16&$-$0.16&$-$0.15\\
		\hline
	\end{tabular}
\end{table}

\subsection{Fixed-binary simulations}
In order to draw a more robust conclusion about the role of Brownian
motion in loss cone refilling, we re-ran the set of simulations with
$\gamma = 1.0$ while periodically re-centering the binary at the
centre of the { merger remnant} to quench the binary's wandering.  To this
purpose, we introduced the following modifications in the $N$-body
integrator: (i) at each integration step, both MBHs were forced to
advance with the smallest populated timestep in the hierarchy; (ii)
the BHB was pinned down at the centre of the density cusp by shifting
the binary CoM position during each corrector step and setting its
velocity to zero. These new runs were started at the time when the
binary separation in the original simulations drops below
$a_{\rm f}/10$. The centre of the stellar density cusp was evaluated by
computing the CoM of particles inside a sphere with radius $l_0$
($\approx 5$), i.e. the sphere containing $50\%$ of stars at the
beginning of these new runs. During each corrector step the predicted
positions of the stars were used to (i) estimate the CoM (S0) of
particles within a distance $l_0$ from the binary CoM and (ii)
reiterate the procedure by computing a more precise CoM (S1) using the
just estimated CoM (S0) as new centre of the sphere.  { We refer to this
new set of integrations as {\it fixed-binary} simulations, to
distinguish them from the previous {\it free-binary} ones.}
\begin{figure*}
  \includegraphics[angle=270,trim={9cm 0 2cm 7cm},width=.7\textwidth]{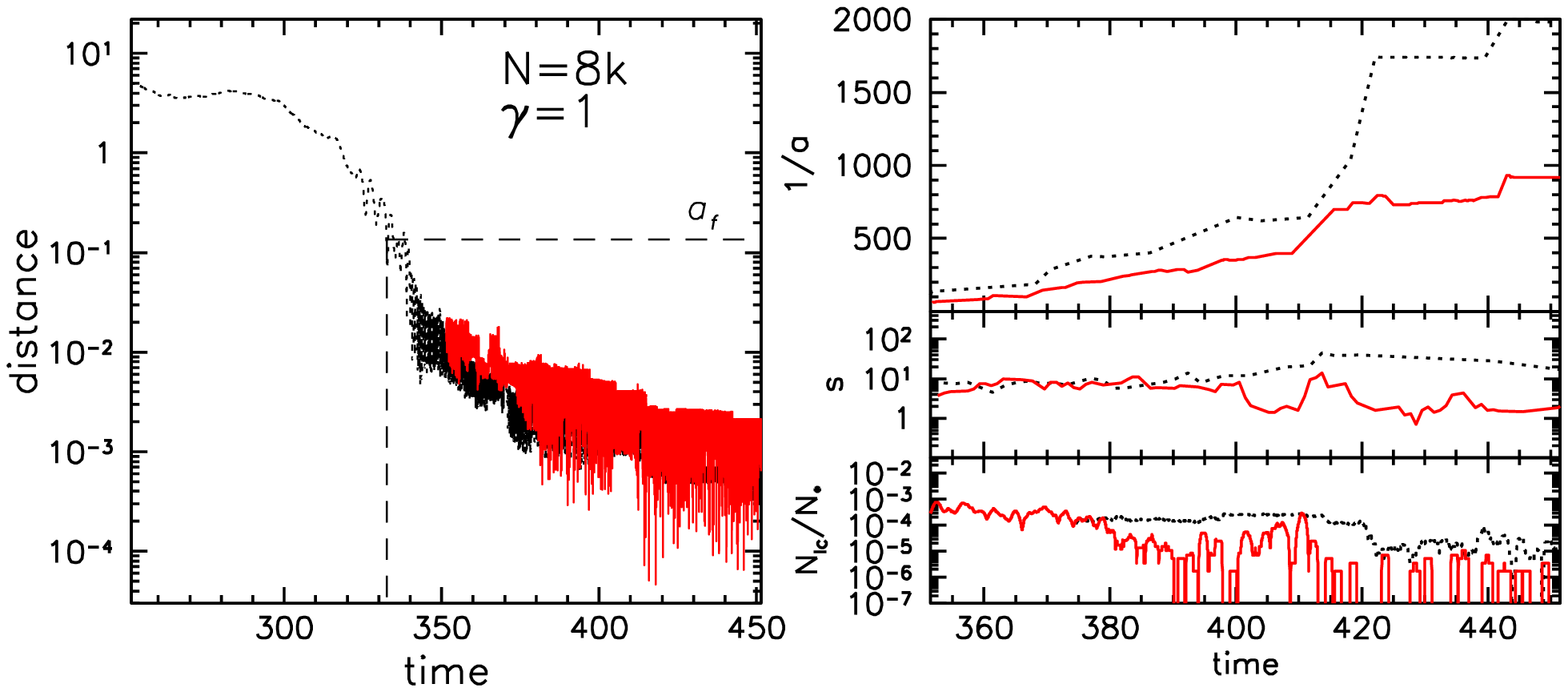}
  \includegraphics[angle=270,trim={9cm 0cm 2cm 7cm},width=.7\textwidth]{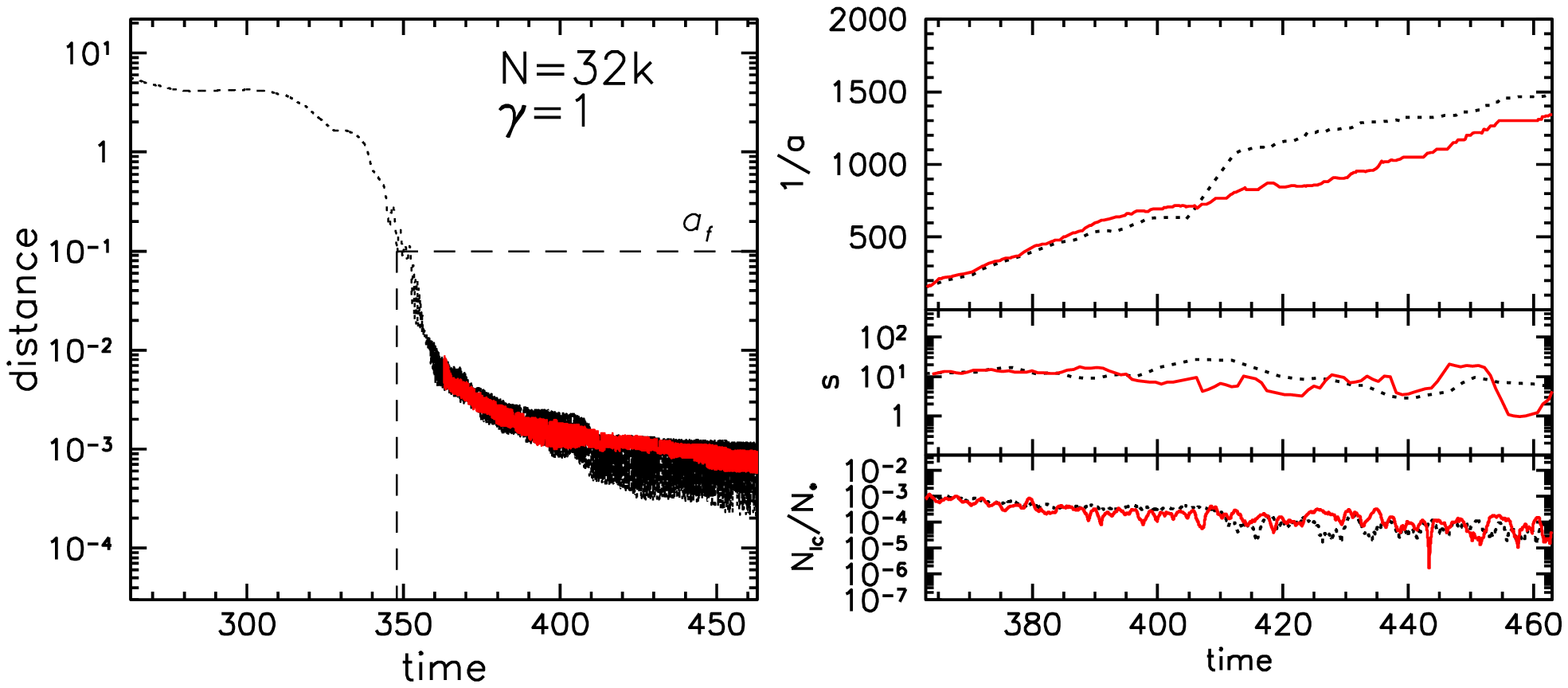} 
  \includegraphics[angle=270,trim={9cm 0 2cm 7cm},width=.7\textwidth]{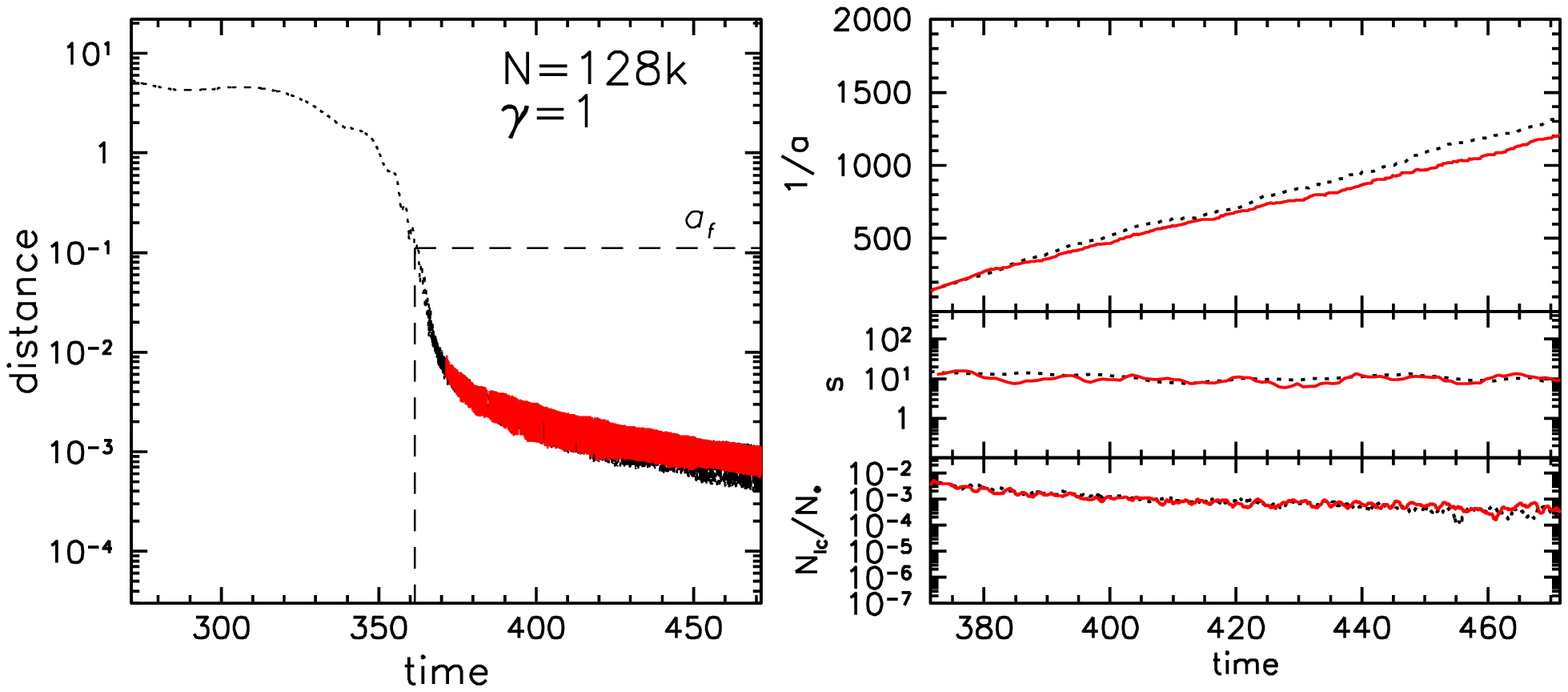}
  \includegraphics[angle=270,trim={9cm 0 2cm 7cm},width=.7\textwidth]{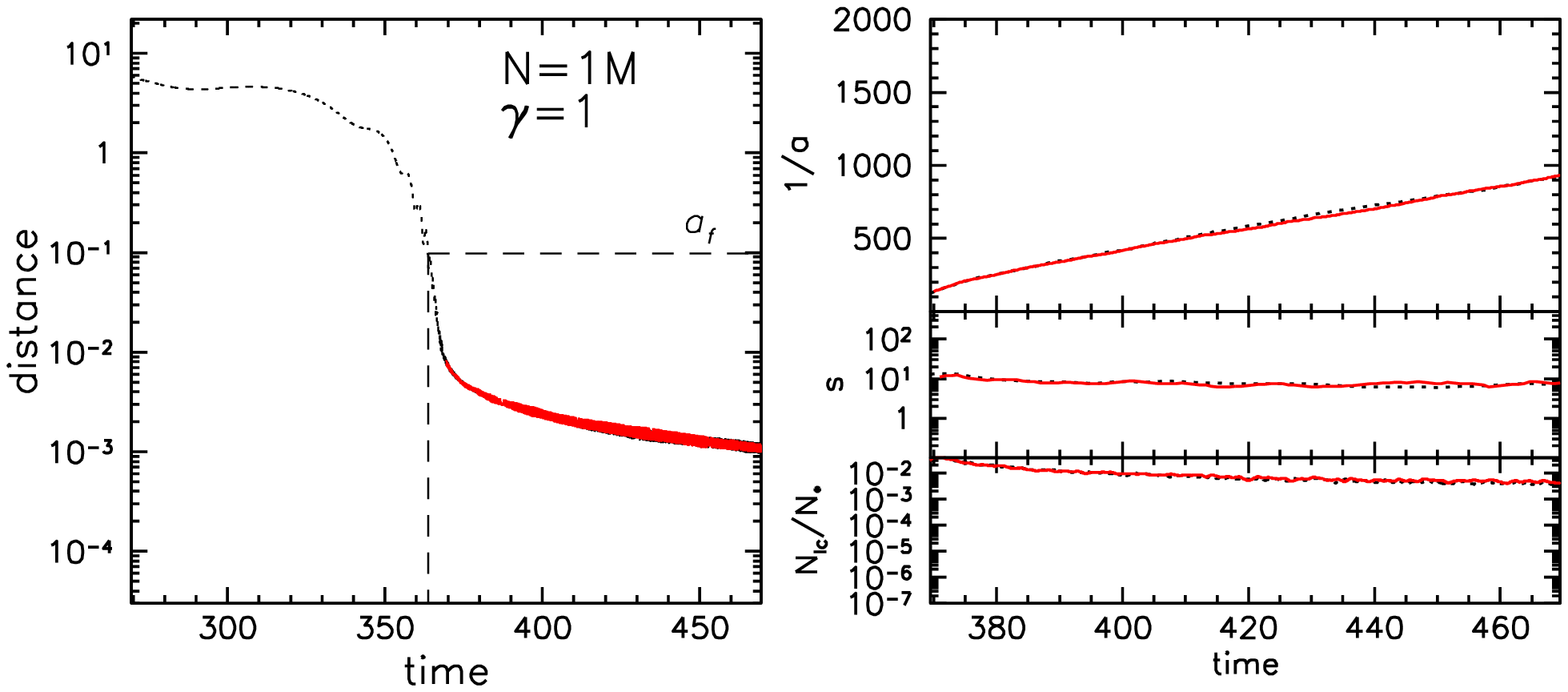}
  \caption{Evolution of the binary separation (left-hand panels) and of the
    inverse semimajor axis $1/a$, hardening rate $s$ and fraction of
    loss cone stars $\nlc/N$ (from top to bottom, right-hand panels). The
    free-binary evolution is shown {with a black dotted line, while the fixed-binary
    evolution is overplotted with a red solid line}. We present the evolution for some
    of the available resolutions: from top to bottom, $N=8$k, 32k,
    128k, 1M. All panels refer to the set of simulations with $\gamma=1$. }
  \label{fig:fix1}
\end{figure*}

\begin{figure}
	\includegraphics[angle=270,trim={9.3cm .0cm 1.9cm 17.3cm},width=.9\columnwidth]{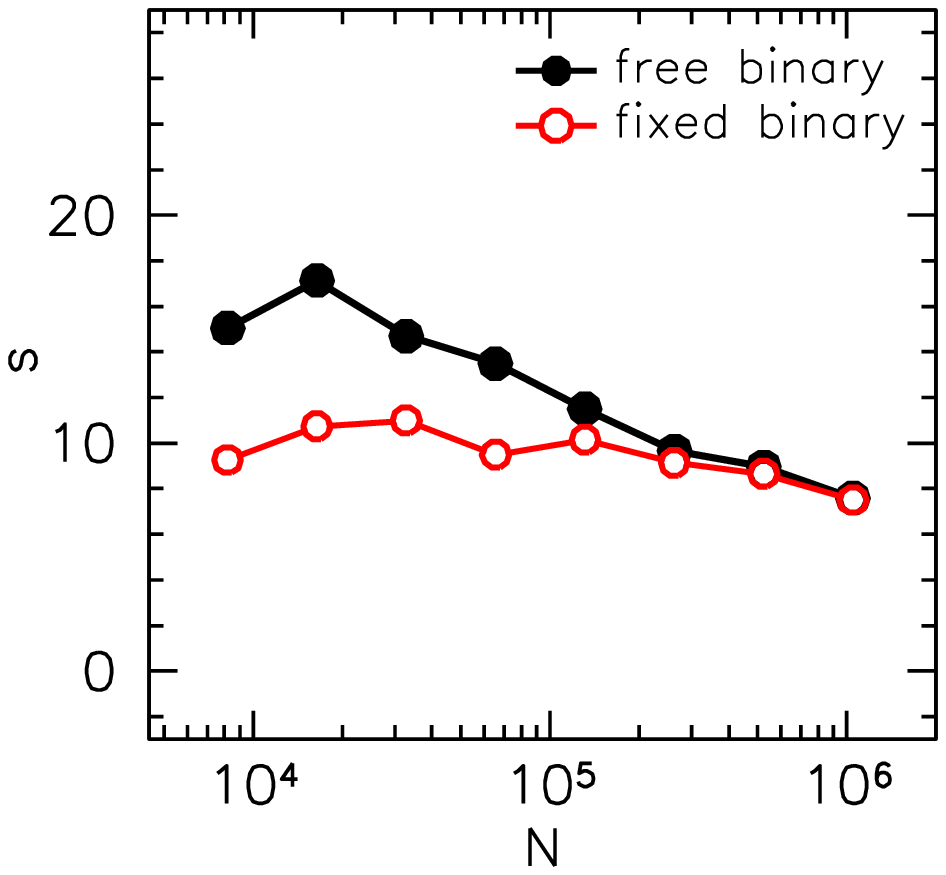}
    \caption{Time-averaged binary hardening rate as a function of the
      number of simulated particles for the $\gamma=1$ set in the case
      of free-binary evolution (black filled circles) and fixed-binary
      evolution (red empty circles). The two sets converge at the
      largest $N$ values. }
    \label{fig:smean}
\end{figure}

{ From Figure \ref{fig:fix1} it is apparent that the hardening rate is lower for
the fixed-binary runs in the low-$N$ limit.}  As the resolution is increased, though, the
free-binary evolution resembles more and more the fixed-binary
evolution. This aspect can be better appreciated from Figure
\ref{fig:smean}: the time-averaged hardening rate in the free-binary
runs approaches the hardening rate of the runs with anchored BHB as
$N$ increases, and the binary behaviour converges in the higher
resolution simulations. Accordingly, the power-law dependence of the
averaged hardening rate on $N$ in the fixed-binary simulations is
shallower ($p_s=-0.052$) compared to the one obtained from the
free-binary runs ($p_s=-0.16$).

In addition, we note that the hardening rate measured in the low-$N$
runs with fixed binary is only slightly larger than the same rate
measured at the highest particle number. This implies that Brownian
motion has a non negligible role in loss cone refilling for particle
numbers in the range $10^4 \leq N \leq 10^6$
{  %
, while it does not appreciably influence the BHB evolution if $N\gtrsim 10^6$.

\citet{chatterjee} and \citet{quinlan1997} used a similar strategy and compared  simulations in which the binary is free to wander with simulations in which the BHB CoM is pinned down at the origin; however they both find that hardening quickly stops after the first loss cone depletion if the binary is not allowed to random walk. 
This discrepancy comes from the fact that (i) \citet{chatterjee} and \citet{quinlan1997} used a different
integration technique with respect to our paper, and  (ii) they simulated the BHB evolution in spherical systems, where triaxial loss cone refilling
is inhibited; this mechanism is indeed expected to be the main driver of the BHB hardening in high-resolution runs.
}

We conclude that Brownian motion has an appreciable influence on the
BHB evolution only at small particle numbers, and it can be safely
neglected in free-binary simulations for $N\gtrsim 1$M.

\section{Summary}
\label{sec:sum}
Massive BHBs are believed to form in the final stages
of galaxy mergers. The formation of MBH pairs in gas-poor environments
is driven by a combination of dynamical friction and close encounters
with stars on low angular momentum orbits. However, both theoretical
models and simulations show that evolution via the slingshot mechanism
may be ineffective in bringing the binaries to coalescence if the loss
cone region is not efficiently replenished. Recent simulations suggest
that a departure from spherical symmetry in the merger remnant may
lead to a change in the main mechanism driving loss cone refilling,
from collisional two-body scatterings to collisionless torques in
non-spherical systems. The reliability of these results may be
questioned given the modest particle number achievable in current
state-of-the-art direct-summation $N$-body simulations.  In such
cases, the inevitable wandering of the binary due to Brownian motion
results in a larger population of stars on loss cone orbits.

In this work, we investigated the significance of Brownian motion of
BHBs in merger simulations in the context of the final parsec
problem. We performed three sets of direct-summation simulations
corresponding to three different choices of the inner slope of the
density profile { ($\gamma{}=0.5,1.0,$ and 1.5), and varying the particle number (from 8k to 1M particles).}

We found that the
effect of Brownian motion on the binary evolution is rather weak, and
the Brownian motion amplitude $r_{\rm b}$ is in good agreement with the
expected $r_{\rm b}\propto N^{-0.5}$ relation, while the binary hardening
rate only exhibits a weaker dependence on $N$.

We also performed
additional simulations in which the binary centre of mass was fixed at
the centre of the density cusp {(fixed-binary runs)}, and we found that the hardening rate
of the BHB measured in the free-binary and fixed-binary case converge
to similar values for $N \sim $1M.  Moreover, the hardening rate measured in low-$N$ models with a fixed-binary evolution is only slightly larger than the one measured in { high-$N$ free-binary runs}. 
This suggests that the role of Brownian motion in merger
simulations is comparable to the effect collisional loss cone
repopulation for $10^4 \lesssim N \lesssim 10^6$.  However, Brownian
motion is not important for $N \gtrsim1$ M even in free-binary
simulations. 

Our findings support the general belief that
non-sphericity of merger remnants leads to efficient collisionless loss
cone refilling \citep{preto,khan11,gualandrismerritt} and the final
parsec problem disappears whenever this dominates over collisional
effects.

\section*{Acknowledgements}
{We warmly thank Monica Colpi and Eugene Vasiliev for useful discussions and suggestions. We also thank the anonymous referee for their useful comments. The $N$-body integrations were carried out on the GPU cards \emph{CUDA GeForce GTX 780}, \emph{CUDA GeForce GTX 780 Ti} at the University
of Surrey and on the \emph{CUDA Tesla K80} GPUs hosted by the Italian
supercomputing consortium \emph{CINECA}. We would like to thank the Astrophysics group at the University of Surrey for the possibility of running most of the presented simulations. We acknowledge the CINECA Award N. HP10CP8A4R, 2016 for the availability of high performance computing resources and support.  EB and MM acknowledge financial support from the Istituto Nazionale di Astrofisica (INAF) through a Cycle 31st PhD grant, from the Italian Ministry of Education, University and Research (MIUR) through grant FIRB 2012 RBFR12PM1F, from INAF through grant PRIN-2014-14, and from the MERAC Foundation.

\bibliography{biblio} 

\begin{thebibliography}{}
\makeatletter
\relax
\def\mn@urlcharsother{\let\do\@makeother \do\$\do\&\do\#\do\^\do\_\do\%\do\~}
\def\mn@doi{\begingroup\mn@urlcharsother \@ifnextchar [ {\mn@doi@}
  {\mn@doi@[]}}
\def\mn@doi@[#1]#2{\def\@tempa{#1}\ifx\@tempa\@empty \href
  {http://dx.doi.org/#2} {doi:#2}\else \href {http://dx.doi.org/#2} {#1}\fi
  \endgroup}
\def\mn@eprint#1#2{\mn@eprint@#1:#2::\@nil}
\def\mn@eprint@arXiv#1{\href {http://arxiv.org/abs/#1} {{\tt arXiv:#1}}}
\def\mn@eprint@dblp#1{\href {http://dblp.uni-trier.de/rec/bibtex/#1.xml}
  {dblp:#1}}
\def\mn@eprint@#1:#2:#3:#4\@nil{\def\@tempa {#1}\def\@tempb {#2}\def\@tempc
  {#3}\ifx \@tempc \@empty \let \@tempc \@tempb \let \@tempb \@tempa \fi \ifx
  \@tempb \@empty \def\@tempb {arXiv}\fi \@ifundefined
  {mn@eprint@\@tempb}{\@tempb:\@tempc}{\expandafter \expandafter \csname
  mn@eprint@\@tempb\endcsname \expandafter{\@tempc}}}

\bibitem[\protect\citeauthoryear{{Aarseth}}{{Aarseth}}{2003}]{aarsethbook}
{Aarseth} S.~J.,  2003, {Gravitational N-Body Simulations}

\bibitem[\protect\citeauthoryear{{Abbott} et~al.,}{{Abbott}
  et~al.}{2016}]{Abbott2016}
{Abbott} B.~P.,  et~al., 2016, \mn@doi [Physical Review Letters]
  {10.1103/PhysRevLett.116.061102}, \href
  {http://adsabs.harvard.edu/abs/2016PhRvL.116f1102A} {116, 061102}

\bibitem[\protect\citeauthoryear{{Babak} et~al.,}{{Babak}
  et~al.}{2016}]{Babak2016}
{Babak} S.,  et~al., 2016, \mn@doi [\mnras] {10.1093/mnras/stv2092}, \href
  {http://adsabs.harvard.edu/abs/2016MNRAS.455.1665B} {455, 1665}

\bibitem[\protect\citeauthoryear{{Begelman}, {Blandford}  \& {Rees}}{{Begelman}
  et~al.}{1980}]{begelman}
{Begelman} M.~C.,  {Blandford} R.~D.,   {Rees} M.~J.,  1980, \mn@doi [\nat]
  {10.1038/287307a0}, \href {http://adsabs.harvard.edu/abs/1980Natur.287..307B}
  {287, 307}

\bibitem[\protect\citeauthoryear{{Berczik}, {Merritt}  \& {Spurzem}}{{Berczik}
  et~al.}{2005}]{BMS2005}
{Berczik} P.,  {Merritt} D.,   {Spurzem} R.,  2005, \mn@doi [\apj]
  {10.1086/491598}, \href {http://adsabs.harvard.edu/abs/2005ApJ...633..680B}
  {633, 680}

\bibitem[\protect\citeauthoryear{{Capuzzo-Dolcetta}, {Spera}  \&
  {Punzo}}{{Capuzzo-Dolcetta} et~al.}{2013}]{higpus}
{Capuzzo-Dolcetta} R.,  {Spera} M.,   {Punzo} D.,  2013, \mn@doi [Journal of
  Computational Physics] {10.1016/j.jcp.2012.11.013}, \href
  {http://adsabs.harvard.edu/abs/2013JCoPh.236..580C} {236, 580}

\bibitem[\protect\citeauthoryear{{Chandrasekhar}}{{Chandrasekhar}}{1942}]{chandra}
{Chandrasekhar} S.,  1942, {Principles of stellar dynamics}

\bibitem[\protect\citeauthoryear{{Chandrasekhar}}{{Chandrasekhar}}{1943}]{dinfric}
{Chandrasekhar} S.,  1943, \mn@doi [\apj] {10.1086/144517}, \href
  {http://adsabs.harvard.edu/abs/1943ApJ....97..255C} {97, 255}

\bibitem[\protect\citeauthoryear{{Chatterjee}, {Hernquist}  \&
  {Loeb}}{{Chatterjee} et~al.}{2003}]{chatterjee}
{Chatterjee} P.,  {Hernquist} L.,   {Loeb} A.,  2003, \mn@doi [\apj]
  {10.1086/375552}, \href {http://adsabs.harvard.edu/abs/2003ApJ...592...32C}
  {592, 32}

\bibitem[\protect\citeauthoryear{{Dehnen}}{{Dehnen}}{1993}]{dehnen}
{Dehnen} W.,  1993, \mnras, \href
  {http://adsabs.harvard.edu/abs/1993MNRAS.265..250D} {265, 250}

\bibitem[\protect\citeauthoryear{{Dorband}, {Hemsendorf}  \&
  {Merritt}}{{Dorband} et~al.}{2003}]{dorband}
{Dorband} E.~N.,  {Hemsendorf} M.,   {Merritt} D.,  2003, \mn@doi [Journal of
  Computational Physics] {10.1016/S0021-9991(02)00067-0}, \href
  {http://adsabs.harvard.edu/abs/2003JCoPh.185..484D} {185, 484}

\bibitem[\protect\citeauthoryear{{Ferrarese} \& {Ford}}{{Ferrarese} \&
  {Ford}}{2005}]{ff05}
{Ferrarese} L.,  {Ford} H.,  2005, \mn@doi [\ssr] {10.1007/s11214-005-3947-6},
  \href {http://adsabs.harvard.edu/abs/2005SSRv..116..523F} {116, 523}

\bibitem[\protect\citeauthoryear{{Gualandris} \& {Merritt}}{{Gualandris} \&
  {Merritt}}{2012}]{gualandrismerritt}
{Gualandris} A.,  {Merritt} D.,  2012, \mn@doi [\apj]
  {10.1088/0004-637X/744/1/74}, \href
  {http://adsabs.harvard.edu/abs/2012ApJ...744...74G} {744, 74}

\bibitem[\protect\citeauthoryear{{Haehnelt} \& {Rees}}{{Haehnelt} \&
  {Rees}}{1993}]{haehneltrees}
{Haehnelt} M.~G.,  {Rees} M.~J.,  1993, \mn@doi [\mnras]
  {10.1093/mnras/263.1.168}, \href
  {http://adsabs.harvard.edu/abs/1993MNRAS.263..168H} {263, 168}

\bibitem[\protect\citeauthoryear{{Hills}}{{Hills}}{1983}]{hills}
{Hills} J.~G.,  1983, \mn@doi [\aj] {10.1086/113418}, \href
  {http://adsabs.harvard.edu/abs/1983AJ.....88.1269H} {88, 1269}

\bibitem[\protect\citeauthoryear{{Khan}, {Just}  \& {Merritt}}{{Khan}
  et~al.}{2011}]{khan11}
{Khan} F.~M.,  {Just} A.,   {Merritt} D.,  2011, \mn@doi [\apj]
  {10.1088/0004-637X/732/2/89}, \href
  {http://adsabs.harvard.edu/abs/2011ApJ...732...89K} {732, 89}

\bibitem[\protect\citeauthoryear{{King}}{{King}}{1966}]{king}
{King} I.~R.,  1966, \mn@doi [\aj] {10.1086/109857}, \href
  {http://adsabs.harvard.edu/abs/1966AJ.....71...64K} {71, 64}

\bibitem[\protect\citeauthoryear{{Makino} \& {Funato}}{{Makino} \&
  {Funato}}{2004}]{makinofunato}
{Makino} J.,  {Funato} Y.,  2004, \mn@doi [\apj] {10.1086/380917}, \href
  {http://adsabs.harvard.edu/abs/2004ApJ...602...93M} {602, 93}

\bibitem[\protect\citeauthoryear{{Merritt}}{{Merritt}}{2001}]{merrittbrown01}
{Merritt} D.,  2001, \mn@doi [\apj] {10.1086/321550}, \href
  {http://adsabs.harvard.edu/abs/2001ApJ...556..245M} {556, 245}

\bibitem[\protect\citeauthoryear{{Merritt}}{{Merritt}}{2005}]{merritt05}
{Merritt} D.,  2005, \mn@doi [\apj] {10.1086/429398}, \href
  {http://adsabs.harvard.edu/abs/2005ApJ...628..673M} {628, 673}

\bibitem[\protect\citeauthoryear{{Merritt}, {Berczik}  \& {Laun}}{{Merritt}
  et~al.}{2007a}]{merritt07}
{Merritt} D.,  {Berczik} P.,   {Laun} F.,  2007a, \mn@doi [\aj]
  {10.1086/510294}, \href {http://adsabs.harvard.edu/abs/2007AJ....133..553M}
  {133, 553}

\bibitem[\protect\citeauthoryear{{Merritt}, {Mikkola}  \& {Szell}}{{Merritt}
  et~al.}{2007b}]{merritt2007}
{Merritt} D.,  {Mikkola} S.,   {Szell} A.,  2007b, \mn@doi [\apj]
  {10.1086/522691}, \href {http://adsabs.harvard.edu/abs/2007ApJ...671...53M}
  {671, 53}

\bibitem[\protect\citeauthoryear{{Milosavljevi{\'c}} \&
  {Merritt}}{{Milosavljevi{\'c}} \& {Merritt}}{2001}]{milosav2}
{Milosavljevi{\'c}} M.,  {Merritt} D.,  2001, \mn@doi [\apj] {10.1086/323830},
  \href {http://adsabs.harvard.edu/abs/2001ApJ...563...34M} {563, 34}

\bibitem[\protect\citeauthoryear{{Milosavljevi{\'c}} \&
  {Merritt}}{{Milosavljevi{\'c}} \& {Merritt}}{2003}]{milosav}
{Milosavljevi{\'c}} M.,  {Merritt} D.,  2003, \mn@doi [\apj] {10.1086/378086},
  \href {http://adsabs.harvard.edu/abs/2003ApJ...596..860M} {596, 860}

\bibitem[\protect\citeauthoryear{{Nitadori} \& {Makino}}{{Nitadori} \&
  {Makino}}{2008}]{nitadorimakino}
{Nitadori} K.,  {Makino} J.,  2008, \mn@doi [\na]
  {10.1016/j.newast.2008.01.010}, \href
  {http://adsabs.harvard.edu/abs/2008NewA...13..498N} {13, 498}

\bibitem[\protect\citeauthoryear{{Plummer}}{{Plummer}}{1911}]{plummer}
{Plummer} H.~C.,  1911, \mnras, \href
  {http://adsabs.harvard.edu/abs/1911MNRAS..71..460P} {71, 460}

\bibitem[\protect\citeauthoryear{{Preto}, {Berentzen}, {Berczik}  \&
  {Spurzem}}{{Preto} et~al.}{2011}]{preto}
{Preto} M.,  {Berentzen} I.,  {Berczik} P.,   {Spurzem} R.,  2011, \mn@doi
  [\apjl] {10.1088/2041-8205/732/2/L26}, \href
  {http://adsabs.harvard.edu/abs/2011ApJ...732L..26P} {732, L26}

\bibitem[\protect\citeauthoryear{{Quinlan} \& {Hernquist}}{{Quinlan} \&
  {Hernquist}}{1997}]{quinlan1997}
{Quinlan} G.~D.,  {Hernquist} L.,  1997, \mn@doi [\na]
  {10.1016/S1384-1076(97)00039-0}, \href
  {http://adsabs.harvard.edu/abs/1997NewA....2..533Q} {2, 533}

\bibitem[\protect\citeauthoryear{{Saslaw}, {Valtonen}  \& {Aarseth}}{{Saslaw}
  et~al.}{1974}]{slingshot}
{Saslaw} W.~C.,  {Valtonen} M.~J.,   {Aarseth} S.~J.,  1974, \mn@doi [\apj]
  {10.1086/152870}, \href {http://adsabs.harvard.edu/abs/1974ApJ...190..253S}
  {190, 253}

\bibitem[\protect\citeauthoryear{{Sesana}}{{Sesana}}{2010}]{sesa10}
{Sesana} A.,  2010, \mn@doi [\apj] {10.1088/0004-637X/719/1/851}, \href
  {http://adsabs.harvard.edu/abs/2010ApJ...719..851S} {719, 851}

\bibitem[\protect\citeauthoryear{{Thorne} \& {Braginskii}}{{Thorne} \&
  {Braginskii}}{1976}]{gw}
{Thorne} K.~S.,  {Braginskii} V.~B.,  1976, \mn@doi [\apjl] {10.1086/182042},
  \href {http://adsabs.harvard.edu/abs/1976ApJ...204L...1T} {204, L1}

\bibitem[\protect\citeauthoryear{{Vasiliev}, {Antonini}  \&
  {Merritt}}{{Vasiliev} et~al.}{2015}]{vasiliev15}
{Vasiliev} E.,  {Antonini} F.,   {Merritt} D.,  2015, \mn@doi [\apj]
  {10.1088/0004-637X/810/1/49}, \href
  {http://adsabs.harvard.edu/abs/2015ApJ...810...49V} {810, 49}

\bibitem[\protect\citeauthoryear{{Volonteri}}{{Volonteri}}{2010}]{volonteri10}
{Volonteri} M.,  2010, \mn@doi [\aapr] {10.1007/s00159-010-0029-x}, \href
  {http://adsabs.harvard.edu/abs/2010A%26ARv..18..279V} {18, 279}

\bibitem[\protect\citeauthoryear{{Yu}}{{Yu}}{2002}]{yu}
{Yu} Q.,  2002, \mn@doi [\mnras] {10.1046/j.1365-8711.2002.05242.x}, \href
  {http://adsabs.harvard.edu/abs/2002MNRAS.331..935Y} {331, 935}

\bibitem[\protect\citeauthoryear{{eLISA Consortium}}{{eLISA
  Consortium}}{2013}]{elisa}
{eLISA Consortium} 2013, preprint, \href
  {http://adsabs.harvard.edu/abs/2013arXiv1305.5720C} {} (\mn@eprint {arXiv}
  {1305.5720})

\makeatother
\end{thebibliography}

\bsp	
\label{lastpage}
\end{document}